\documentclass[useAMS,usenatbib,usegraphicx]{mn2e}

\usepackage{epsfig}
\usepackage{amssymb}
\usepackage{myaasmacros}

\def \m{\mathrm}

\title[Inside-out growth of ellipticals]{How do minor mergers promote inside-out growth 
of ellipticals, transforming the size, density profile and 
dark matter fraction?} 
\author[Hilz et al.]{Michael Hilz$^{1}$,Thorsten Naab$^{1}$\thanks{E-mail:
naab@mpa-garching.mpg.de} and J.P. Ostriker$^{2}$\\
$^{1}$Max-Planck-Institut f\"ur Astrophysik, Karl-Schwarzschild-Str. 1, 85741 Garching, Germany\\
$^{2}$Department of Astrophysical Sciences, Princeton University, Princeton, NJ 08544, USA}

\begin{document}

\date{Accepted ???. Received ??? in original form ???}

\pagerange{\pageref{firstpage}--\pageref{lastpage}} \pubyear{2008}

\maketitle

\label{firstpage}

\begin{abstract}
There is observational evidence for inside-out growth of giant
elliptical galaxies since $z \gtrsim 2-3$, which is - in contrast to
disk galaxies - not driven by in-situ star formation. Many of the $\sim  
10^{11} M_{\odot}$ systems at high redshift have small sizes
$\sim 1kpc$ and surface brightness profiles with low Sersic indices
n. The most likely descendants at $z = 0$ have, on average, grown by a
factor of two in mass and a factor of four in size, indicating $r
\propto M^{\alpha}$ with $\alpha \gtrsim 2$. They also have surface
brightness profiles with $n \gtrsim 5$. This evolution can be
qualitatively explained on the basis of two assumptions: compact
ellipticals predominantly grow by collisionless minor (mass-ratio 1:10) or
intermediate (mass-ratio 1:5) 'dry' mergers,
and they are embedded in massive dark matter halos which support the
stripping of merging satellite stars at large radii. We draw these 
conclusions from idealized collisionless mergers spheroidal galaxies -
with and without dark  matter - with mass ratios of 1:1, 1:5, and
1:10. The sizes evolve as $r \propto M^{\alpha}$ with $\alpha < 2$ for
mass-ratios of 1:1 (and 1:5 without dark matter halos) and , while doubling
the stellar mass, the Sersic index increases from $n \sim 4$ to $n
\sim 5$. For minor mergers of galaxies embedded in dark matter halos,
the sizes grow significantly faster and the profile shapes change more
rapidly. Surprisingly, already mergers with moderate mass-ratios of
1:5, well motivated by recent cosmological simulations, give $\alpha
\sim 2.3$ and after only two merger generations ($\sim 40$ per
cent added stellar mass) the Sersic index has increased to $n
> 8$ ($n \sim 5.5$ without dark matter), reaching a final value of
$n = 9.5$ after doubling the stellar mass. This is accompanied by a
significant increase of the dark matter fraction (from $\sim 40$ per cent to 
$\gtrsim 70$ per cent) within the stellar half-mass radius, driven by the strong size 
increase probing larger, dark matter dominated regions. For equal-mass
mergers the effect is much weaker. We conclude that only a few
intermediate mass-ratio mergers ($\sim 3-5$ with initial mass-ratios
of 1:5) of galaxies embedded in massive dark matter halos can result in the
observed concurrent inside-out growth and the rapid evolution in
profile shapes. This process might explain the existence of
present day giant ellipticals with sizes, $r > 4 kpc $, high Sersic indices,
$n  > 5$, and a significant amount of  dark matter within the
half-light radius. Apart from negative stellar metallicity gradients
such a 'minor' merger scenario 
also predicts significantly lower dark matter fractions for $z \sim 2$
compact quiescent galaxies and their rare present day analogues.      
\end{abstract}


\section{INTRODUCTION}
\label{intro}

Merging is a natural process in hierarchical cosmological models and
plays a significant role for the assembly of massive galaxies
(e.g. \citealp{1996MNRAS.281..487K}). The most massive objects are
elliptical galaxies, which are considered to start forming their stars
at a redshift of $z\sim 6$ in a dissipative environment and can rapidly
become very massive ($\sim 10^{11}M_{\odot}$) by $z =2$
\citep{2005MNRAS.363....2K,2006ApJ...648L..21K,2006MNRAS.366..499D,2006ApJ...649L..71K,2007ApJ...658..710N,2009ApJ...699L.178N,2009ApJ...692L...1J,2009ApJ...703..785D,2009MNRAS.395..160K,2010ApJ...725.2312O,2010ApJ...709..218F,2011MNRAS.417..900D,2011ApJ...736...88F,2012ApJ...744...63O}.  
Their subsequent evolution is not fully understood yet, as these
ellipticals are observed to be already quiescent at $z \sim 2$, on
average 4-5 times  smaller, and a factor of two less
massive than their low redshift descendants 
\citep{2005ApJ...626..680D,2006ApJ...650...18T,2007MNRAS.374..614L, 
2007ApJ...671..285T,2007ApJ...656...66Z,2007MNRAS.382..109T,2007ApJ...656...66Z,  
2008ApJ...687L..61B,2008ApJ...677L...5V,2008A&A...482...21C,2008ApJ...688..770F,
2009MNRAS.392..718S,2009ApJ...695..101D,2009ApJ...700..221K,2009ApJ...697.1290B,
2010ApJ...709.1018V,2010A&A...524A...6S,2012ApJ...745..179W,2012arXiv1206.1540O}. This
rapid size evolution eventually proceeds in an inside-out fashion in
the absence of significant star formation an is accompanied by an
evolution in the light (and eventually mass) distribution 
\citep{2009MNRAS.398..898H,2010ApJ...709.1018V,2011MNRAS.411L...6A,2011MNRAS.411.1435T,2012ApJ...749..121S,2012MNRAS.422.3107S}. This
growth process is distinctively different from the star formation
driven inside-out growth of disk galaxies
\citep{1995A&A...302...69P,1989MNRAS.239..885M,1998MNRAS.295..319M,2006MNRAS.366..899N,2007MNRAS.374.1479G,2011MNRAS.412.1081W,2012MNRAS.422..997K}. In
addition, there is
growing observational evidence that high redshift ellipticals are more  
flattened with exponential like surface brightness distributions ($n \lesssim 4$)
whereas their potential present day massive descendants are rounder
and have more concentrated profiles shapes with $n \gtrsim 5$ 
  \citep{2005ApJ...624L...9T,2006ApJ...650...18T,2007ApJ...671..285T,2008ApJ...677L...5V,2008ApJ...688...48V,2009ApJ...698.1232V,2010ApJ...709.1018V,2011ApJ...730...38V,2009ApJS..182..216K,2010MNRAS.405.2253C,2011ApJ...743...87W,2011arXiv1111.6993B,2012ApJS..198....2K,2012ApJ...749..121S}. 

Direct observations of massive compact and quiescent high redshift
systems also provide new constraints on the formation histories of
giant elliptical galaxies. At face value the dramatic difference in
properties  of massive quiescent galaxies at high and low redshift
rules out a simple 'monolithic' formation 
followed by simple passive evolution \citep{2008ApJ...682..896K,2008ApJ...677L...5V}. The observations also
rule out the formation of ellipticals in a single 'disk merger'
event, a scenario for massive ellipticals which might also suffer from
other problems
\citep{1980ComAp...8..177O,2003ApJ...597..893N,2008ApJ...685..897B,2009ApJ...690.1452N}. Even
if the progenitor disks were gas-rich leading  
to a compact remnant \citep{2010MNRAS.406..230R,2010ApJ...722.1666W,2011ApJ...730....4B}
those would have to evolve further by a separate process (see
e.g. \citealp{2009MNRAS.398..898H}). 

In addition to the well studied stellar components there is evidence
for the existence for dark matter within the half-light radii of ellipticals
\citep{2001AJ....121.1936G,2009ApJ...691..770T,2011MNRAS.415..545T,2012arXiv1202.3308C}. Gravitational  
lensing measurements of massive galaxies in the local universe
predict, on average, $\sim$ 30 per cent of the matter within the half-light
radius of present day massive elliptical galaxies being 
dark matter \citep{2010ApJ...724..511A,2011MNRAS.415.2215B}. At high redshift,
however, the situation is more uncertain \citep{2012arXiv1204.3099T}
as dark matter cannot be measured directly. The naive expectation is
that dark matter is much less important due to the dissipative nature
of the early evolution. Still, assuming its existence and
collisionless nature, dark matter will eventually  affect the
distribution of stars during the further assembly of the galaxies (see
e.g. \citealp{2010ApJ...712...88L,2012arXiv1206.1597H}).  

The underlying idealized assumption for the study presented here is
that present day massive elliptical galaxies ($> 10^{11}M_{\odot}$)
since $z \sim 2$ have grown predominantly by accreting stars that have
formed in other galaxies and this process can be approximated by
simulations of mergers of collisionless 'dry' stellar systems. This
assumption is not too far fetched as there is clear observational
evidence for the predominance of old ($z \gtrsim 2$) stellar populations
in these galaxies, leaving little room for gas accretion and
subsequent star formation
(e.g. \citealp{1973ApJ...179..427S,2000ApJ...536L..77B,2005ApJ...622L...5T,2010MNRAS.404.1775T,2010ApJ...709.1018V,2010ApJ...710.1170L,2011MNRAS.414..940Y,2012MNRAS.421.1298C}).  
Directly observed mergers, inferred merger rates, as well as stellar
fine structure at large galactocentric radii are indicative of stellar
merger and accretion events
\citep{1983ApJ...274..534M,1992AJ....104.1039S,2005ApJ...627L..25T,2005AJ....130.2647V,2006ApJ...640..241B,2006ApJ...652..270B,2007ApJ...665..265F,2008MNRAS.388.1537M,2008ApJ...676L.105W,2010ApJ...719..844R,2011MNRAS.417..863D,2011ApJ...742..103L,2011MNRAS.415.3903T,2011ApJ...743...87W,2012ApJ...746..138T,2012arXiv1202.4674L}.   
On the theoretical side, phenomenological as well as direct numerical
studies predict that the late assembly of massive galaxies is
dominated by the late accretion of stars that formed early
\citep{1996MNRAS.281..487K,2006MNRAS.366..499D,2007ApJ...658...65C,2007ApJ...658..710N,2008MNRAS.384....2G,2009ApJ...699L.178N,2011arXiv1110.1420Y,2010ApJ...712...88L,2010ApJ...725.2312O,2011MNRAS.413..101G,2011ApJ...736...88F,2012ApJ...744...63O,2012ApJ...752...41Y,2012arXiv1202.5315G,2012arXiv1205.5807M,2012arXiv1206.0295L}).  
In a previous study \citet{2012arXiv1206.1597H} used idealized merger simulations
of spheroidal galaxies to re-investigate the effect of the galaxy merger 
mass-ratio - limited to the extreme cases of 1:1 and 1:10 - on the
merger driven evolution of compact high-redshift spheroids. The main
finding of this study was that both minor and major mergers lead to
size growth and an increase of the dark matter fraction, however by
different physical processes and at significantly different
strengths. Violent relaxation in major mergers mixes dark matter into
the central regions and escaping particles limit the expected size
growth even below the expected values. In 1:10 mergers on the other
hand, satellite particles are stripped at large radii where the initial
host galaxies are dominated by dark matter. As a result the stellar
effective radii and the dark matter fractions of the galaxies grow rapidly.

Here we extend the \citet{2012arXiv1206.1597H} study to focus on the
inside-out growth including a moderate mass-ratio of 1:5. This step is
important for a number of reasons: The average stellar mass growth
since $z \sim 2$ is a factor of two. Mergers with mass-ratios of
1:10 typically take a long time to complete and it is not clear
whether ten of these mergers can even be completed in a Hubble 
time. Additionally, recent observationally estimated merger rates indicate
more minor mergers for massive galaxies and at higher redshift but the
rates are still low and it is not clear - also theoretically - whether there are enough minor
mergers to explain the size growth
\citep{2009ApJ...703.1531N,2009ApJ...706L..86N,2011ApJ...738L..25W,2011ApJ...743...87W,2012ApJ...746..162N,2012MNRAS.422.1714N,2012MNRAS.422L..62C,2012ApJ...744...85M,2012arXiv1206.1612E,2012MNRAS.422.2187M}.     
Higher mass-ratio mergers result in a more rapid mass growth and
fewer of them would be needed provided the size grows rapid
enough. Another reason to focus on mass-ratios of 1:5 was recently 
provided by direct cosmological simulations. Three independent 
numerical studies have found that the average mass-weighted mass ratio 
of galaxy mergers building massive ellipticals is 1:4 - 1:5, making
this regime particularly interesting
\citep{2012ApJ...744...63O,2012arXiv1202.5315G,2012arXiv1206.0295L}.  

In section \ref{sims} we give a short review of the initial galaxy
models and the simulation parameters. In the subsequent Sections
\ref{sizeev}, \ref{surfd}, \ref{secfit}, and \ref{dark2}  we
investigate the evolution of sizes, surface densities, profiles shapes
and dark matter fractions, respectively. We discuss the results and conclude in
Section \ref{summary}.  

\section{SIMULATIONS}
\label{sims}

\begin{figure}
  \begin{center}
    \includegraphics[width=8.5cm]{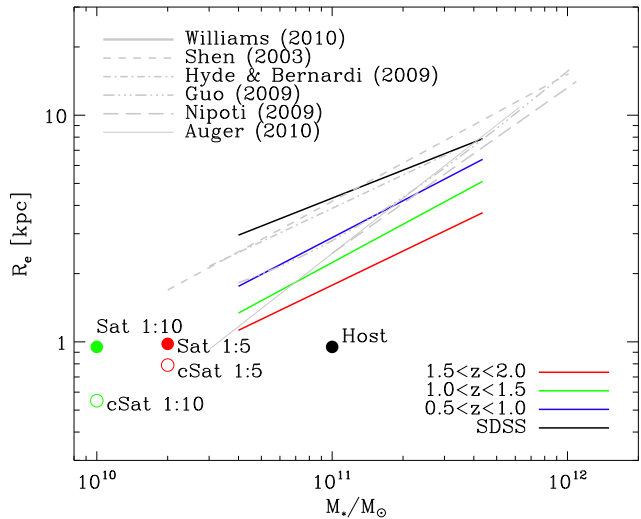}
  \end{center}
  \caption[title]
          {
            Stellar mass versus effective radius for the initial
            galaxy models. For our fiducial set of simulations
            (diffuse satellites) all galaxies
            have an effective radius of $1kpc$ (filled circles). For the
            second set of simulations with compact satellites (open
            circles) the sizes follow the $z=2$ relation (red line) by
            \citet{2010ApJ...713..738W}. The initial conditions are
            well separated from local mass-size relations of ellipticals
            \citep{2003MNRAS.343..978S,2009MNRAS.394.1978H,2009MNRAS.398.1129G,2009ApJ...706L..86N,2010ApJ...724..511A}.
          }
  \label{fig1}
\end{figure}

Continuing the long line of research in this direction
\citep{1978MNRAS.184..185W,1979MNRAS.189..831W,1980ApJ...235..421M,1983MNRAS.204..219V,1983ApJ...265..597F,2003MNRAS.342..501N,2005MNRAS.362..184B,2007ApJ...658...65C,2009ApJ...703.1531N}
we present a set of simulations of dissipationless mergers of spheroidal  
galaxies with and without dark matter halos and with mass ratios of
1:1, 1:5, and 10:1. Details about the initial models and the 
simulations parameters for 1:1 and 1:10 mergers are presented in
\citet{2012arXiv1206.1597H}. The simulations were performed with the
N-body/SPH code $VINE$
\citep{2009ApJS..184..298W,2009ApJS..184..326N}. For the present study
we have also performed comparison run with $GADGET$
\citep{2005MNRAS.364.1105S} with identical results. Here we only give
a brief summary of the  
simulation setup. The initial galaxy models are isotropic, spherically
symmetric one- and two-component systems following Hernquist density
profiles \citep{1990ApJ...356..359H} either for a model representing
only the stellar component of the galaxy  
(one-component, bulge-only) or a bulge embedded in a massive dark matter 
halo (two-component, bulge+halo). The host bulges have a stellar mass of
$M_{\m{*,host}}=1$ and a scale radius of $a_{\m{*,host}}=1.0$ (corresponding to
a projected half-mass radius of $R_{\mathrm{e}} \sim 1.8 a_{\m{*,host}}$
(\citealp{1990ApJ...356..359H}) realized with 100.000 particles. For
the two-component cases we assume an additional dark matter halo with
a total dark to stellar mass ratio of $M_{\m{dm}}/M_{*}=10$ and a ratio of the
scale radii of $a_{\m{dm}}/a_{*}=11$. The halo is realized with 1.000.000
particles resulting in equal-mass particles for both components. The
gravitational softening length is set to $\epsilon =  
0.02$ for all particles. The stability of the initial N-body models was
demonstrated in \citet{2012arXiv1206.1597H}. The merging satellite galaxies are 5
(mass-ratio 1:5) or 10 times (mass-ratio 1:10) less massive than the
host and have a correspondingly lower particle number. For our
fiducial set of simulations we assume the same scale radius, $a_{\m{*,sat}}=1.0$, for
all bulges, resulting in satellites being more diffuse than the
host. For a second set of simulations we chose more compact satellites
with scale radii of $a_{\m{*,sat}}=0.8$ for the 1:5 mergers and
$a_{\m{*,sat}}=0.5$ for the 1:10 mergers following a
$z \sim 2$ mass-size relation \citep{2010ApJ...713..738W}. 

The open and filled circles in Fig. \ref{fig1} indicate the location
of the initial galaxy models in 
the mass-size plane compared to several published present day and high
redshift mass-size relations. The initial massive 
host galaxies resemble typical massive and compact 'red nuggets' at
$z=2$ which are smaller than typical present day galaxies at $10^{11}
M_{\odot}$ (red line from \citealp{2010ApJ...713..738W}).  However,
for redshift two galaxies below $\sim 4.0 \times 10^{10}M_{\odot}$
(which would be the minor merger partners) there is no  
consistent size information available. Therefore we assume two scenarios: a fixed size
of $\sim 1kpc$ for the satellites, similar to local galaxies (see
e.g. \citealp{2011MNRAS.414.3699M}), indicated by the filled red and green
circles. Here the merging lower mass galaxies are diffuse spheroids
with low phase space densities comparable to low mass disk-like
galaxies. Alternatively we use smaller sizes consistent with a simple
continuation of the $z =2$ mass-size relation to lower masses (open
circles), resulting in compact spheroids.  

The first generation of equal-mass mergers are parabolic mergers of one-
and two-component models of the initial host galaxies. The second generation
is a re-merger of the duplicated, randomly oriented, first generation 
merger remnant, which was allowed to dynamically relax at the
center. The randomly oriented galaxies approach each other on
parabolic orbits with a pericenter distance of half the spherical
half-mass radius of the progenitor remnants, i.e. the pericenter
distances increase with each merger generation. The sequences of
intermediate and minor mergers with initial mass-ratios of 1:5 (1:10)
are also simulated with 
one- and two-component models. In the following we use the traditional
convention to call all mergers with mass-ratios smaller than 1:4
minor mergers. Initially, the mass-ratio is 1:5 (1:10)
and the galaxies are set on parabolic orbits. The randomly
oriented merger remnants of the first generations are then set on
parabolic orbits with the initial satellite galaxy models and a 
mass-ratio of now 1:6 (1:11), and so on. We perform 6 generations of 1:10
mergers and 5 generations of 1:5 mergers using satellites with
fixed scale radii (Sat 1:5/1:10, see Fig. \ref{fig1}). For
comparison, we also performed minor mergers of two-component models with
compact satellites (cSat 1:5/1:10, see Fig. \ref{fig1}). Again all
parabolic orbits have pericenter distances of half the spherical 
half-mass radius of the bulge of the massive progenitor galaxy.

The 1:10 bulge-only simulations have no dark matter halo and the
in-falling satellites suffer less from dynamical friction. Therefore
the final coalescence takes by far the longest time. However, taking
about $\sim 9$Gyrs for 10 merger generations even this process could 
be completed by the present day assuming a $z=2$ progenitor. All other 
merger series are completed in less than $\sim 7$Gyrs. 

\begin{figure}
  \begin{center}
    \includegraphics[width=8.5cm]{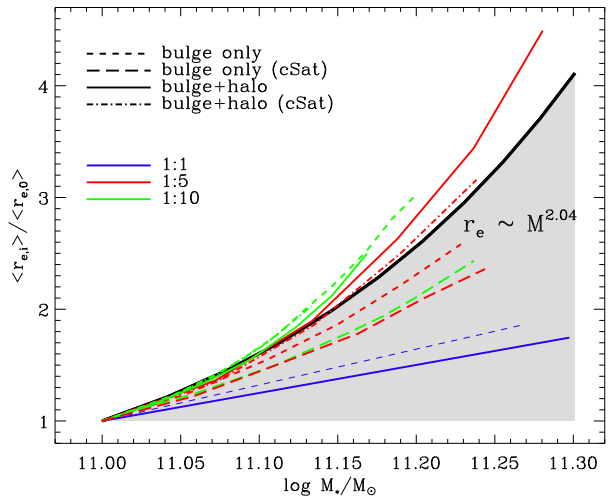}
  \end{center}
  \caption[title]
          {
            The projected spherical half-mass radius of the stellar
            component (the mean value along the three principal axes)
            as a function of bound stellar mass for 1:1 (blue), 1:5 (red), and 1:10
            (green) mergers. The observed size growth is indicated by
            the solid black line (see \citet{2010ApJ...709.1018V} for
            the observational uncertainties). The size
            evolution of models in the grey shaded area is too
            weak to be consistent with observations. All mergers of bulges
            embedded in massive dark matter halos and high mass-ratios 
            (1:5, 1:10, red and green solid/dashed-dotted lines) show a rapid
            size evolution. The size evolution of the
            bulge-only models (short and long dashed lines) are not efficient
            enough, except the 1:10 scenario with a diffuse satellite (green dashed line).
            The accretion of compact satellites results in less size
            growth compared to the diffuse satellites. The major
            merger lines (blue) show size growth that is far too
            slow. 
          }
  \label{fig2}
\end{figure}

\section{Evolution of the sizes}
\label{sizeev}

After the completion of every merger, we allow the central region of
the remnant to relax, before we compute the projected circular
half-mass radii, $r_{e}$, along the three principal axes and the
bound stellar (bulge) mass, $M_*$. 

In Fig. \ref{fig2} we show the
evolution of the half-mass radius as a function of the bound stellar
mass for 1:1 (blue), 1:5 (red), and 1:10 (green) merger
hierarchies. The black line 
indicates the observed evolution, $r_{e} \propto$M$^{2.04}$, in the
mass-size plane from $z \sim 2$ to the present day
\citep{2010ApJ...709.1018V}. The models here are idealized in the
sense that they do not contain any dissipative component. If present,
like in the 'real' universe, this component would reduce the size
growth per added mass \citep{2006MNRAS.370.1445D}. Therefore we consider a model a failure if it
occupies the the shaded area in Fig. \ref{fig1}. Promising models have
to lie above this line so that small amounts of gas - and therefore
somewhat smaller sizes - can be tolerated.  

Equal-mass mergers show an almost linear increase of size with mass,
(see also
\citealp{2005MNRAS.362..184B,2007ApJ...658...65C,2009ApJ...697.1290B,2009ApJ...703.1531N,2012arXiv1206.1597H}), 
independent of whether the stellar system is embedded in a dark matter
halo or not (blue solid and dashed lines). As discussed by \citet{2005MNRAS.362..184B}, 
in mergers with dark matter halos the in-falling galaxy suffers more
from dynamical friction in the massive dark matter halo of the
companion galaxy, resulting in more energy transfer from the bulge to
the halo, leading to a more tightly bound bulge with a smaller size
(blue solid line, Fig. \ref{fig2})  compared to the model without dark
matter (blue dashed line, Fig. \ref{fig2}).  If we combine the results
of both major merger scenarios this yields a mass-size relation of
$r_{e}\propto $M$^{0.91}$ which is comparable to the results of
\citet{2005MNRAS.362..184B}, who found a slightly smaller exponent  ($\sim
0.7$) for orbits with high angular momentum and an exponent $\sim 1$
for pure radial orbits. Nevertheless, as the size grows at most
linearly with mass, dissipationless major mergers cannot be the main
driver for the size evolution of early-type galaxies. 

As expected from simple virial estimates
\citep{2000MNRAS.319..168C,2007ApJ...658...65C,2009ApJ...699L.178N,2009ApJ...697.1290B},
the size evolution is 
stronger for bulge-only models with lower mass-ratios of 1:5 and
1:10 (red and green lines in Fig. \ref{fig2}). However,
except for the 1:10 mergers with a diffuse satellite (green  
short dashed line), all minor mergers with bulge-only satellites are
not efficient enough to escape the 'forbidden' area. This behaviour is
improved for minor mergers of two-component models, where bulges are
embedded in a massive dark matter halos. For mass-ratios 1:5 and
1:10 the size evolution is in excess of the observed evolution.  In
the case of 1:5 minor mergers with a less compact satellite (red
solid line), we obtain a mass-size growth relation of $r_{e}\propto
$M$^{2.4}$ with a similar or even larger exponent for both
two-component 1:10 scenarios. Therefore we consider all minor merger
models (1:5 and 1:10) with dark matter halos and the diffuse 1:10
mergers to be consistent with observations even in more realistic
models, where dissipational effects would reduce the size growth 
\citep{2006ApJ...641...21R,2006ApJ...650..791C,2008ApJ...689...17H,2011MNRAS.415.3135C}.  
The size growth via major mergers is too slow to fit observational
data. 

\begin{figure*}
  \begin{center}
    \includegraphics[width=18cm]{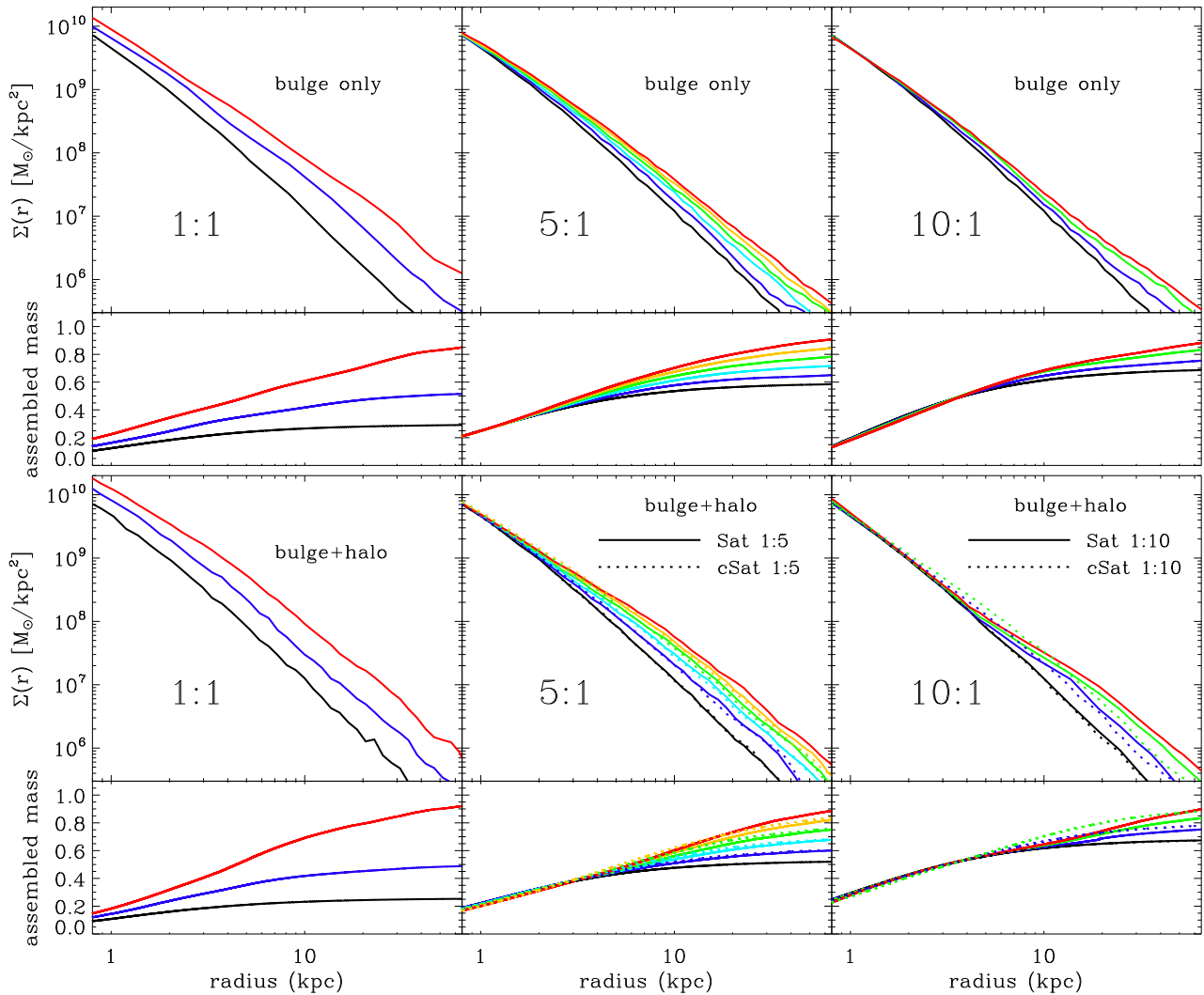}
  \end{center}
  \caption[title]
          {
            Top panels: Stellar surface mass density profiles and
            cumulative stellar mass
            distributions for the fiducial 1:1 (left), 1:5 (middle) and 1:10
            (right) bulge-only merger models after each merger event
            (colored lines from bottom to top, for 1:1 and 1:5 mergers
            we show every generation, for 1:10 mergers every second).  
            For all mass-ratios the surface densities increase at all
            radii with merger generation with a trend of a stronger  
            increase in the outer regions for lower mass-ratios.   
            Bottom panels: Same for mergers of systems with dark matter
            halos. Equal-mass mergers show a similar behaviour than
            bulge-only models. For 1:5 and 1:10 bulge+halo models
            (mergers with compact satellites are indicated by the
            dotted lines) the inside-out
            growth trend is stronger than for bulge-only
            models. Regions inside 4kpc are almost unaffected
            and the satellite stellar mass assembles predominantly at
            large radii similar to observations
            \citet{2010ApJ...709.1018V}.              
          }
  \label{fig3}
\end{figure*}

\section{Evolution of the surface densities}
\label{surfd}

We now take a closer look at the evolution of the surface densities of
the merger remnants. In Fig. \ref{fig3}, we plot
the stellar surface mass densities and cumulative stellar mass
profiles after every 1:1 and 1:5 merger and every second 1:10
merger for bulge-only models (top panels) and bulge+halo models
(bottom panels). For equal-mass mergers there is mass growth at all radii,
i.e. the lines are shifted more or less parallel to higher densities
independent of the presence of a halo (see
e.g. \citealp{1980ApJ...235..421M,1983MNRAS.204..219V,1983ApJ...265..597F}
for discussions on the weak but existing break of homology). This
evolution scenario would be in disagreement to observations of \citet{2010ApJ...709.1018V},
which show, that early-type galaxies grow inside-out, i.e. the central
densities stay constant and most of the mass  assembles at larger
radii, building up an extended envelope of stars.  
\begin{figure*}
  \begin{center}
    \includegraphics[width=18cm]{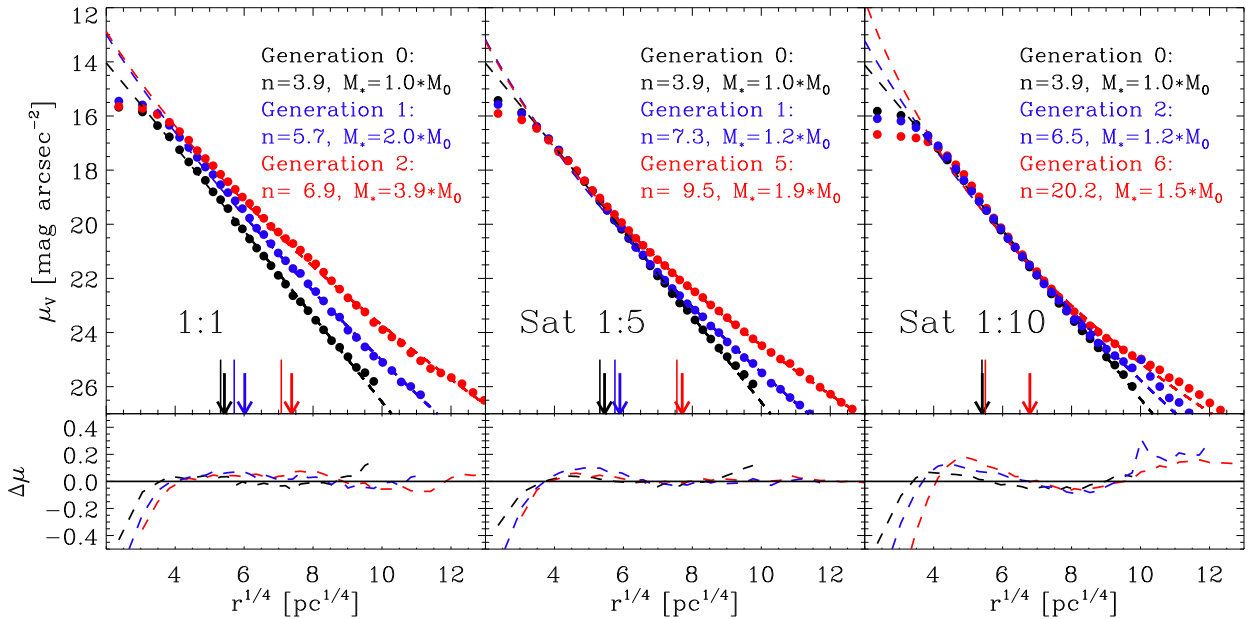}
  \end{center}
  \caption[title]
          {Surface brightness profiles $\mu_{\mathrm{V}} (r)$ of the fiducial
            two-component 1:1 (left), 1:5 (middle), and 1:10 (right)
            merger remnants as a function of radius for the initial
            conditions (black circles), the first merger generation
            (blue circles, for 1:10 we show the second generation) and
            the final remnants (red circles). The over-plotted dashed
            lines show the best fitting Sersic function for the outer
            profiles. The data was fitted from $0.02 r_{\mathrm{e}}$ to either
            $10 r_{\mathrm{e}}$ or a limiting surface brightness of
            $m_{V}=27\mathrm{mag/arcsec}^2$. The 
            residuals $\Delta \mu < 0.2 \mathrm{mag/arcsec}^2$ (lower
            panels) are small, except for the 1:10 case, which cannot
            be reasonably well fitted anymore for late generations ($n
            > 20$)   The fitted 
            effective radii $r_{\mathrm{e,fit}}$ (short vertical lines) are slightly
            smaller than the projected half-mass radii $r_{\mathrm{e}}$ (arrows). 
            For minor mergers - 1:5 and 1:10 - the best fitting Sersic
            index $n$ increases rapidly as the mass added in minor
            mergers grows. Note that for minor mergers the central
            surface densities decline the most even as the outer
            surface densities (and Sersic indices) increase.  
          }
  \label{fig4}
\end{figure*}

The second column in Fig. \ref{fig3} depicts the surface densities and
mass assembly of minor mergers with an initial mass ratio of
1:5. For the bulge-only models (top) with a diffuse satellite  
(Sat 1:5, Fig. \ref{fig2}), the surface density stays nearly constant
up to $r \sim 1-2$kpc and increases mainly in the outer parts. This
effect was already reported by \citet{1983MNRAS.204..219V} as a
possible explanation for abundance gradients in elliptical galaxies. The
same scenario leads to an even stronger inside-out growth if the
galaxies are surrounded by a dark matter halo (bottom panel,
second column). Here the bulge particles get stripped at larger radii
and the central surface density ($ r < 4kpc$) stays unaffected; it only 
increases at radii $r > 2-3$kpc. The size growth shown in
Fig. \ref{fig2} is due to this build-up  of a massive stellar
envelope. The mass is added to large radii where, prior to the
merger, the dark matter component was large. Due to our accounting
procedure the dark matter fraction within $r_{\m{e}}$ increases, even
though the dark matter added in such mergers to the inner parts is
negligible.  The dotted lines in these panels show the four remnants, 
where the satellites are more compact (cSat 1:5, Fig. \ref{fig2}) and
lie an extension of the $z\sim 2$ mass-size relation. Obviously, the results are
very similar.

The six generations of minor mergers with an initial mass-ratio of
1:10 are shown in the right column of Fig. \ref{fig3}. In the case of
bulge-only models (top panels), the surface density increases predominantly at larger radii,
similar to the previous scenario, but now the satellite stars are even
less bound compared to the 1:5 case and therefore are stripped at
larger radii, even without a dark matter halo. If present, this effect 
is enhanced (lower panels of right column), as the satellites first
orbit through the massive dark matter halo - and lose stars -  before
they reach the center of the host. The surface densities are unaffected inside $r=5$kpc. 
In summary, 1:5 and 1:10 mergers lead to a significant change in the
mass distribution of the galaxies with most of the stellar satellite material
assembling at large radii. This picture hardly changes for compact
satellites (cSat, dotted lines), although the scale lengths of the
satellite stars are significantly smaller, they are more bound and
resists the drag force of the host potential for a longer
time. Consequently, somewhat more material gets closer to the central regions. 
In summary a minor merger driven evolution scenario - in particular
for 1:5 bulge+halo mergers - is in good agreement with the picture we
get from observations. Although we are definitely affected by
poor numerical resolution at the very centers of the galaxies we note
the differential effect that for minor mergers the central surface
densities decline the most even as the  outer surface densities (and
Sersic indices) increase in agreement with expectations from recent observations
\citep{2012ApJ...749..121S,2012ApJ...751...45T}.

\section{Evolution of profile shapes}
\label{secfit}

The curvature of the light profiles of elliptical galaxies is an
important parameter as it correlates with other observed properties of
elliptical galaxies, such as the effective radius $r_{e}$, the total luminosity and the stellar mass 
\citep{1993MNRAS.265.1013C,2003MNRAS.342..501N,2006MNRAS.369..625N,
2009ApJS..182..216K}. We therefore use a Sersic $r^{1/n}$ \citep{1968adga.book.....S} 
function to fit synthetic surface brightness profiles of our
simulations, 
\begin{eqnarray}
I(r) = I_{e}\cdot 10^{-b_{n}((r/r_{e})^{1/n}-1)},
\label{sersic}
\end{eqnarray}
where the three free parameters are the effective surface brightness $I_{e}$, the effective 
radius $r_e$ and the so called Sersic index $n$. Here, $n=1$
corresponds to an exponential profile and $n=4$ to the familiar de
Vaucouleurs profile \citep{1948AnAp...11..247D}. 
The factor $b_n$, is chosen such that the effective radius 
$r_e$ encloses half of the total luminosity. For the expected range
of Sersic indices, this factor can be approximated by the relation $b_n=0.868n-0.142$ 
\citep{1993MNRAS.265.1013C}. We convert the projected surface mass
densities discussed in section \ref{sizeev} to a V-band surface
brightness profile assuming a stellar age of $10^{10}$yr and solar metallicity
$Z=0.02$ \citep{2003MNRAS.344.1000B} which is then fitted with a
Sersic function (see \citet{2006MNRAS.369..625N} for details of the
fitting procedure). Those population properties are reasonable for present day
massive ellipticals \citep{2010MNRAS.404.1775T} but not necessarily so for higher
redshifts. However, details of the choice of population age and
abundance are unimportant for the determination of the profile shape
(which are the same as for the surface density profiles) as  
long as we assume a constant mass-to-light ratio (for the stellar
component) with radius. 


Figure \ref{fig4} shows examples of the Sersic fits to the surface
brightness profiles and the residuals of our fiducial bulge+halo
mergers with mass-ratios of 1:1 (left), 1:5 (middle) and 1:10 (right).
The profiles are fitted from $0.02 r_{e}$ to either $10 r_{e}$ or to a
limiting surface brightness of $m_{V} = 27$ 
mag/arcsec$^{-2}$ \citep{2004AJ....127.1917T,2009ApJS..182..216K}. The
residuals are very small ($\Delta \mu < 0.2$mag/arcsec$^2$), except
for the innermost regions, where the profiles are affected by
insufficient sampling, softening and relaxation effects. The initial
Hernquist spheres  (black circles in all panels) have a Sersic index
of $n=3.9$ (see also \citealt{2006MNRAS.369..625N}). In all cases the fitted Sersic
profiles overestimate the central surface brightness leading to a
slightly smaller fitted effective radius $r_{\m{e,fit}}$  (narrow vertical lines
at the bottom of each surface brightness panel) than the projected
half-mass radius $r_{e}$ (corresponding arrows) presented in
Fig. \ref{fig2}. The differences are small and do not affect any
conclusions in the paper. 

In the case of 1:1 mergers of two-component models (left panel, Fig \ref{fig4}), 
we can see that the profile shape hardly changes for the remnants. The Sersic 
index increases weakly from $n=3.9$ to $n=5.5$ with a corresponding
increase of mass by a factor of about two, hardly enough to
explain the very high numbers observers find for large massive elliptical 
galaxies ($n\sim 10$, see \citealt{1993MNRAS.265.1013C,2009ApJS..182..216K}). 
This picture changes dramatically for minor mergers where the Sersic
index increases rapidly to values of $n \sim 10$ for a similar
increase in mass. This is a direct consequence of the different evolution of
the surface density profiles discussed in section \ref{surfd}. 

\begin{figure}
  \begin{center}
    \includegraphics[width=8.5cm]{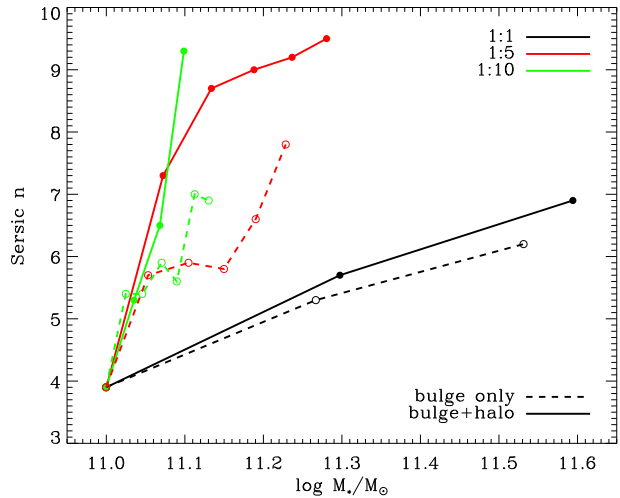}
  \end{center}
  \caption[title]
          {Evolution of the Sersic indices for all fiducial major and
            minor merger remnants shown in Fig. \ref{fig4} as a function of
            bound stellar mass (the evolution for compact satellites
            is similar). All equal-mass mergers show a very
            weak evolution (black lines). Lower mass-ratios lead to a
            stronger evolution for 1:5 (red dashed) and 1:10 (green
            dashed) bulge-only models. The presence of a dark matter
            halos even enhances this effect as more stars are
            deposited at large radii (red and green solid
            lines).  The 1:5 two-component model shows the strongest
            evolution reaching values of $n > 9$ after mass increase of
            only 50 per cent. For 1:10 mergers we only show the first
            generations with reasonable Sersic fits. Later generations
            cannot be accurately fitted anymore (see Fig. \ref{fig4}).
          }
  \label{fig5}
\end{figure}

In Fig. \ref{fig5} we show, for all fiducial bulge-only and bulge+halo
models, the evolution of the Sersic index as a function of the assembled
bound stellar mass. There is only a moderate increase in
Sersic index for the equal-mass mergers. In contrast, after two
generations of 1:5 mergers with a mass increase of of only 40 per cent
the Sersic index can be $n>7$ (for the bulge+halo model) and the final
remnant reaches values of $n \sim 9.5$, which is in the range of the observed
present day massive elliptical galaxies \citep{1993MNRAS.265.1013C,2009ApJS..182..216K}.  
The corresponding bulge-only minor merger scenarios (red and green
dashed lines in Fig. \ref{fig5}) show similar but weaker trends yielding final
Sersic  indices of $n \sim 7.5$. Still, the overall evolution is
much faster for two-component models. This again indicates that the
merger mass-ratio and dark matter halos play an important role as they
increase the effect of stripping at large radii in a way, that the accreted
stellar mass assembles at the  'right' regions of the host 
galaxy.

\section{Evolution of dark matter fractions}
\label{dark2}
\begin{figure}
  \begin{center}
    \includegraphics[width=8.5cm]{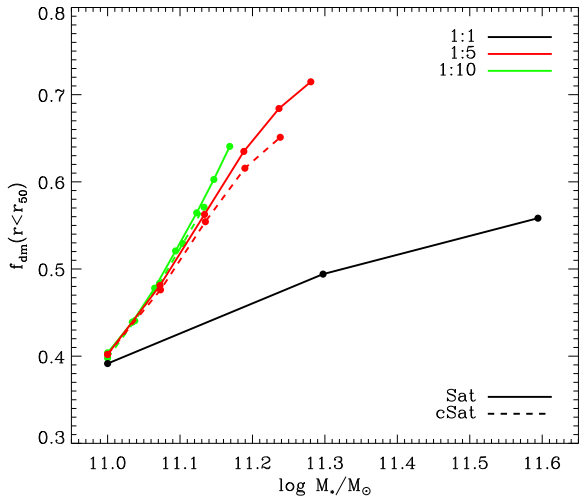}
  \end{center}
  \caption[title]
          {Evolution of the dark matter fraction,
            $f_{\m{dm}}(r<r_{\mathrm{50}})=M_{\mathrm{dm}}(r<r_{\mathrm{50}})/M_{\mathrm{tot}}(r<r_{\mathrm{50}})$,
            within the spherical half-mass radius $r_{\m{50}}$ (which
            is similar to the half mass radius) for all bulge+halo
            models as a function of bound stellar mass of the merger
            remnants. Equal-mass mergers show a weak increase of
            $f_{\m{dm}}$. A very rapid increase of $f_{\m{dm}}$ is
            found for 1:5 (red) and 1:10 (green) mergers, with a
            similar evolution for compact satellites (dashed lines).
            The evolution of $f_{\m{dm}}$ for 1:5 mergers with mass is
            very similar to 1:10 mergers indicating that less events
            results in similar change of $f_{\m{dm}}$.   
          }
  \label{fig7}
\end{figure}

In this section we investigate the dark matter fractions within the
effective radii of our simulated merger remnants in the light of recent
lensing observations, which predict an increasing dark matter fraction
for more massive ellipticals
\citep{2010ApJ...724..511A,2011MNRAS.415.2215B}, dynamical modeling
\citep{2001AJ....121.1936G,2012arXiv1202.3308C} and the possibly low dark matter 
fractions in high redshift galaxies \citep{2012arXiv1204.3099T}. The dark 
matter fractions $f_{dm}$ for all bulge+halo simulations 
\begin{eqnarray}
f_{dm}(r<r_{50})=M_{dm}(r<r_{50})/M_{tot}(r<r_{50})
\end{eqnarray}
are shown in Fig. \ref{fig7} as a function of assembled stellar
mass. Here $r_{50}$ denotes the spherical half-mass radius of the
stellar component and $M_{dm}$, $M_{tot}$ are the halo mass and total
mass within $r_{50}$. We have already explained how the observable
$f_{\m{dm}}$ will increase simply by accounting, when the stars are
added far out in the dark matter halo. The dark matter fraction 
increases rapidly with each subsequent minor merger generation
regardless of the mass-ratio. For a mass increase of a factor of two the dark matter
fractions increases by almost a factor two from $\sim 40$ per cent to
$\gtrsim  70$ per cent for minor mergers whereas equal-mass mergers
only an increase to $\sim 55$ per cent.  The strong evolution 
with added mass for minor mergers is a consequence of the rapid  size growth
(Fig. \ref{fig2}), which is in good agreement with
\citep{2009ApJ...703.1531N}. The evolution of $f_{dm}$ correlates with
the radii of the merger remnants. Therefore the 1:5 scenario with 
more compact satellites (red dashed line), which shows slightly weaker
size growth than the fiducial case (Fig. \ref{fig2}), also has a
slightly lower dark matter fraction. On the other hand, the 1:10
bulge+halo remnants grow rapidly in size and therefore have the
highest dark matter fractions. It is easy to understand this trend. As
noted, the dark matter is not pushed inwards; rather we are adding
stars to the outer, dark matter dominated parts of galaxies so
naturally the  amount of dark matter within the stars increases. A
detailed analysis of this process can be found in \citet{2012arXiv1206.1597H}.  
 
\section{Discussion and Conclusion}
\label{summary}
We present idealized binary merger simulations of spheroidal
galaxies with initial mass-ratios of 1:1, 1:5, and 1:10 represented by
pure bulges and bulges embedded in dark matter halos. The 1:5
mass-ratio is of particular interest as it is similar to the average
mass weighted galaxy merger mass-ratio driving the assembly of massive
galaxies recently published cosmological simulations and therefore
those minor mergers (or intermediate mass-ratio mergers,
traditionally all mergers with mass ratios smaller than 1:4 are called
minor mergers) represent a typical growth mode
\citep{2012ApJ...744...63O,2012arXiv1202.5315G,2012arXiv1206.0295L}. 
With this study we aim at assessing the effect of pure collisionless
major and minor mergers on the structural evolution for an already
existent, observed, high-redshift population of compact and massive
galaxies (see e.g. \citet{2007A&A...476.1179B} for a discussion on the
effect of repeated minor mergers on disks). 

Confirming earlier studies we find
that major mergers alone cannot explain the observed size growth but minor 
mergers with mass-ratios of 1:5 or 1:10 show a significantly stronger
size growth per added stellar mass
\citep{2006ApJ...648L..21K,2009ApJ...699L.178N,2009ApJ...706L..86N,2010MNRAS.401.1099H,2012arXiv1206.1597H,2012ApJ...744...63O}. This
effect is slightly enhanced for more diffuse satellites and
significantly enhanced if the galaxies are surrounded by dark matter
halos. In these cases the satellite stars are more efficiently
stripped at larger radii, either because they are less bound or
because they orbit in the deeper potential wells of the dark matter
halos. The latter effect (see \citet{2012arXiv1206.1597H} for a
detailed discussion) leads to an inside-out size growth of $r \propto
M^{2.3}$ already for mergers with a mass-ratio of 1:5, in agreement
with observations  \citep{2010ApJ...709.1018V,2012ApJ...749..121S}.
Overall, the picture is very similar to the dynamical friction driven
galactic cannibalism described by \citet{1978ApJ...224..320H} as an
explanation for BCG properties
\citep{2009ApJ...696.1094R,2012arXiv1202.2357L}.  

The inside-out assembly of mass at larger radii results
in a significant change of the surface density profiles which is here
quantified by an increase of the Sersic index $n$. Starting at a 
fiducial value of $n=4$ the increase is weak for major mergers but
very strong for minor mergers of galaxies embedded in dark matter
halos. In the most extreme case only two 1:5 mergers of bulges with
dark matter halos change the Sersic index from $n \sim 4$ to $n
\sim 8.5$ at a concurrent stellar mass increase of only 40 per
cent. We note that $n=4$ is the fiducial starting point of the
specific simulation setup we have chosen. Qualitatively a similar 
increase in Sersic index is expected for smaller, and probably more
realistic, starting values of $n \sim 2$ for quiescent high redshift
galaxies. In this case the maximum Sersic index might not reach as
high values as presented here but more accurate predictions are
beyond the scope of idealized collisionless studies like the one presented here. 
Major merger scenarios with disk-like progenitors do also
result in an increase of the Sersic index but accompanied by a very
weak size increase or even a reduction in size in the presence of gas 
\citep{2006MNRAS.369..625N,2006ApJ...641...21R,2006ApJ...650..791C,2008ApJ...689...17H,2010ApJ...722.1666W}. Here
the increase in Sersic index is partly driven by dissipative processes
at the center of the remnants. Consistent with the results
presented here, even considering subsequent 'dry' major mergers keep
the Sersic indices for the main stellar body low at $n \lesssim
5$. \citep{2006MNRAS.369..625N,2009ApJ...691.1424H}. 
Therefore a minor merger driven scenario - the late addition of old stars at large radii
- provides a powerful explanation for the predominance of high Sersic
indices of present day large and massive ellipticals
\citep{2001AJ....122.1707G,2004AJ....127.1917T,2009ApJS..182..216K,2011MNRAS.411.2439H,2012ApJS..198....2K}
some of which are most likely the descendants of compact 'red and dead' flattened
high-redshift galaxies with lower Sersic indices 
\citep{2008ApJ...677L...5V,2009ApJ...706L.120V,2010ApJ...719.1969B,2010ApJ...722.1666W,2011arXiv1111.6993B,2011ApJ...742...96W,2011ApJ...730...38V,2012ApJ...749..121S}.  

Naturally, major mergers are also expected to happen during the assembly of
massive galaxies. There is direct and indirect observational evidence
for this but the expected rates are low ($\sim 0.5 - 2$ 
since $z \sim 2$)
\citep{2006ApJ...640..241B,2008MNRAS.388.1537M,2009ApJ...706L.120V,2010ApJ...719..844R,2011ApJ...738L..25W,2012ApJ...746..162N,2012ApJ...744...85M,2012arXiv1202.4674L}.
A major merger will definitely dominate the late mass
assembly history of the galaxy with a significant impact on its
abundance gradients, kinematics and, eventually, morphology  
\citep{1978MNRAS.184..185W,2006ApJ...636L..81N,2011MNRAS.416.1654B}. However, as we have
shown here, the expected evolution in size, surface density profile
shapes and, eventually, dark matter fraction will be only moderate and
not sufficient to evolve the compact high-redshift population into 
present day ellipticals. Another argument against major mergers driving
the size evolution is the apparent absence (or very low number) of massive
compact galaxies in the present day universe
\citep{2009ApJ...692L.118T,2010ApJ...720..723T}. Due to the low
rates a  significant number of massive galaxies will not have
experienced any major merger since $z \sim 2$ and would remain compact
\citep{2009ApJ...697.1290B}. Nevertheless, a few massive compact have
been found which eventually are relics of the high redshift population
or have formed recently in a similar manner 
\citep{2010ApJ...721L..19V,2012ApJ...749L..10J,2012ApJ...751...45T,2012MNRAS.423..632F}. 
Consistent with our framework these galaxies have low Sersic indices
and our model predicts lower dark matter fractions compared to normal sized 
present day giant elliptical galaxies of similar mass. 

The evolution of dark matter fractions within the observable stellar
half-mass radius is also significantly different for major and minor mergers. In
the equal-mass mergers presented here the dark matter fraction increases
from the fiducial starting value of $\sim 40$ per cent to $\sim 48$
per cent for a stellar mass increase of a factor of two. This increase
is driven by mixing processes during the 
violent merger process which lead to a real change in the radial distribution
of luminous and dark matter (\citealp{2012arXiv1206.1597H}, see also
\citealp{2005MNRAS.362..184B}). For the same increase in mass, minor
mergers (e.g. 1:5) result in a four times stronger increase of the
dark matter fraction (from $\sim 40$ per cent to $\gtrsim 70$ per
cent). Here the host galaxy structure is only weakly affected but the
effective radius of the 
stellar distribution increases significantly into regions which were
ab-inito dominated by dark matter. Therefore only the ruler with which the
galaxies are measured is changing 
\citep{2009ApJ...691.1424H,2009ApJ...696L..43C,2010ApJ...712...88L,2012arXiv1206.1597H}. 
We note that the exact numbers for the final dark matter fractions are
determined by our choice of the initial dark matter fraction. Our
fiducial starting value of $\sim 40$ per cent might in reality be
lower for massive high-redshift galaxies but also in this case we
expect similar trends.  There is evidence for the presence of dark
matter in present day massive ellipticals, 
however the exact amount is uncertain 
\citep{2001AJ....121.1936G,2009ApJ...691..770T,2011MNRAS.415..545T,2012arXiv1202.3308C}. 
Most recent lensing observations indicate dark matter fractions (with
large uncertainties) in the range of 20-50 per cent with higher
fractions for more massive galaxies
\citep{2010ApJ...724..511A,2011MNRAS.415.2215B}. These trends are easy
to reconcile in the context of an evolutionary scenario where the
early formation of ellipticals is dominated by a dissipative formation
with little dark matter at the center (see \citep{2012arXiv1204.3099T}
for tentative observational evidence) followed by a dissipationless
assembly dominated by minor mergers naturally increasing the dark
matter fractions. This increase would be larger for more massive
galaxies which assemble more mass by late stellar mergers (see
e.g. \citealp{2006MNRAS.366..499D,2007MNRAS.375....2D,2008MNRAS.384....2G,
2010ApJ...725.2312O,2012ApJ...744...63O,2012arXiv1206.0295L}). However,
predicted dark matter fractions depend sensitively on assumptions about
the stellar initial mass function and recent evidence for rising
stellar mass-to-light ratios with stellar mass leave less room for the
presence of dark matter
\citep{2010Natur.468..940V,2012Natur.484..485C,2012arXiv1205.6473C,2012arXiv1205.6471V,2012arXiv1206.1594F}.  

More quantitative predictions, some based on simple binary merger simulations
of spheroids similar to the ones presented here, in a cosmological
context have been attempted and come to varying conclusions; some
agree with observations \citep{2007ApJ...658...65C}, some highlight possible tension 
with observational trends
\citep{2009ApJ...703.1531N,2009ApJ...706L..86N,2012MNRAS.422L..62C,2012MNRAS.422.1714N,2012ApJ...752L..19Q}. 
However, the quantitative predictive power of such studies is limited
by construction as a realistic cosmological assembly of massive galaxies is very
complex and many details depend on the model assumptions about the
structure of the galaxies and the satellites, 
galaxy orbits, gas fractions, merger rates etc. Qualitatively,
however, our results are in good agreement with high-resolution
cosmological simulations
\citep{2010ApJ...709..218F,2012ApJ...744...63O,2012arXiv1206.0295L},
which are supposedly the method of choice to evaluate these processes
in more detail in the future. 

The predictions of this simple - dark matter assisted - minor merger
driven evolution model are in good qualitative agreement with careful
observations of the structural redshift evolution of early-type galaxies
\citep{2010ApJ...709.1018V,2011ApJ...743...87W,2011arXiv1111.6993B,2012ApJ...749..121S,2012arXiv1205.4058M}. 
We note here that we have tested for the extreme case of compact
progenitors. A significant fraction of high redshift ellipticals have
larger sizes \citep{2010MNRAS.401..933M,2012ApJ...749..121S} and would require
relatively little evolution and less mergers. All the above effects  are
expected to be stronger for more massive galaxies which are expected
to have a larger fraction of their stars acquired in accretion
events. The most extreme cases would be centrals in galaxy clusters 
\citep{1977ApJ...217L.125O,1998ApJ...502..141D,2007MNRAS.375....2D,2009ApJ...696.1094R,2012arXiv1202.2357L}. 

If we are right, then certain definite and testable predictions can be
made about the properties of the outer parts ($r > r_{\mathrm{e}}$) of normal giant
ellipticals: Since they are the result of accretion of low mass
systems they should have considerably lower metallicities than the
inner parts. Steeper gradients at large radii then might imply a
formation history dominated by minor mergers
(\citealp{1978MNRAS.184..185W,2004MNRAS.347..740K}, see however
\citealp{2010MNRAS.407.1347P} for an alternative view). The stellar orbits
at large radii should become radially biased
\citep{2012arXiv1206.1597H} and the dark matter fractions of massive
ellipticals should increase rapidly at large radii. 

\section*{Acknowledgments}

We thank the anonymous referee for valuable comments on the
manuscript. We also thank Simon White for valuable discussions. This
research was supported by the DFG cluster of excellence 'Origin and
Structure of the Universe' and the DFG priority program SPP 1177.

\label{lastpage}

\end{document}